%% file: sigir2020demo.tex
\documentclass[sigconf]{acmart}
\usepackage{multicol}
\usepackage{graphicx}
\usepackage{url}
\usepackage{wrapfig}
\usepackage{slashbox}
\usepackage{balance}
\usepackage{cleveref}
\usepackage{multirow}
\usepackage{subfigure}
\usepackage[export]{adjustbox}
\newcommand{\system}{{{\textsf{BRENDA}}}}
\newcommand{\politifact}{{{\textsf{PolitiFact}}}}
\newcommand{\snopes}{{{\textsf{Snopes}}}}
\newcommand{\declare}{{{\textsf{DeClarE}}}}

\newcommand{\sadhan}{{{\textsf{SADHAN}}}}

\newcommand{\dhan}{{{\textsf{DHAN}}}}

\begin{document}
\begin{abstract}Misinformation such as fake news has drawn a lot of attention in recent years. It has serious consequences on society, politics and economy. This has lead to a rise of manually fact-checking websites such as Snopes and Politifact. However, the scale of misinformation limits their ability for verification. In this demonstration, we propose \system~a browser extension which can be used to automate the entire process of credibility assessments of false claims. Behind the scenes \system~uses a tested deep neural network architecture to automatically identify fact check worthy claims and classifies as well as presents the result along with evidence to the user. Since \system~is a browser extension, it facilities fast automated fact checking for the end user without having to leave the Webpage.

\end{abstract}



\copyrightyear{2020} 
\acmYear{2020} 
\setcopyright{acmlicensed}\acmConference[SIGIR '20]{Proceedings of the 43rd International ACM SIGIR Conference on Research and Development in Information Retrieval}{July 25--30, 2020}{Virtual Event, China}
\acmBooktitle{Proceedings of the 43rd International ACM SIGIR Conference on Research and Development in Information Retrieval (SIGIR '20), July 25--30, 2020, Virtual Event, China}
\acmPrice{15.00}
\acmDOI{10.1145/3397271.3401396}
\acmISBN{978-1-4503-8016-4/20/07}

\title{\system: Browser Extension for Fake News Detection}

\author{Bjarte Botnevik}
\affiliation{University of Stavanger, Norway}
\email{b.botnevik@stud.uis.no}

\author{Eirik Sakariassen}
\affiliation{University of Stavanger, Norway}
\email{e.sakariassen@stud.uis.no}

\author{Vinay Setty}
\affiliation{University of Stavanger, Norway}
\email{vsetty@acm.org}

\maketitle 


\input{introduction}
\input{relatedwork}

\input{system.tex}

\input{demo.tex}

\section{Conclusion}
In this demonstration we proposed \system~which is a browser extension to tackle the challenge of misinformation. The user can use \system~to first identify fact check worthy claims in any news article online. Subsequently the user gets the credibility classification using a sophisticated deep neural network model. The users are also presented with the evidence from the model, and can achieve all this without leaving the Web page of the news article they are reading.

\balance
\bibliographystyle{splncs04}
\bibliography{references}

\end{document}

%% file: introduction.tex
\section{Introduction}
\label{sec:intro}

Online fake news has become a major societal challenge due to its consequences in real life. For example, there are instances of stock market disruptions \footnote{http://business.time.com/2013/04/24/how-does-one-fake-tweet-cause-a-stock-market-crash/}, election meddling \footnote{\url{https://www.theguardian.com/commentisfree/2016/nov/14/fake-news-donald-trump-election-alt-right-social-media-tech-companies}} and mob lynchings \footnote{\url{https://en.wikipedia.org/wiki/Indian_WhatsApp_lynchings}}. To address this, several fact checking organizations such as Snopes, Politifact and FullFact have become popular. Typically they employ experts and journalists who perform a tedious task of manually selecting fact check worthy claims made in online news and social media debunking them.

We propose \system~a proof of concept browser extension which anyone can install on desktop browsers to perform end-to-end fact checking. \system~automates following two tasks: (1) Selecting fact check worthy claims and (2) Verifying the truthfulness of claims based on the evidence found online. 

Existing demos (e.g, CredEye \cite{Popat:2018:CCL:3184558.3186967},  FactMata\footnote{https://try.factmata.com/}, \cite{Miranda:2019:AFC:3308558.3314135} etc) are limiting to the users reading online news, since they have to first identify the claims within the articles, then switch to a different website for fact checking. There are also demos which either do only claim ranking \cite{claimrank} or just list the relevant websites  \cite{zhi2017claimverif}.  Moreover, existing demos do not provide any explanation for the claim classifications. There are no existing demos which can jointly identify the claim and fact check them and provide evidence to the support the decision. To address these issues, \system~provides the following contributions:

\begin{enumerate}
    \item \system~facilitates users to do fact checking without leaving the Website. If users are not sure on which claims to fact check, \system~can automatically filter fact check worthy claims.
    \item \system~can automatically query online evidence via web search engines and verify claims.
    \item \system~uses a proven pre-trained deep neural network model coined~\sadhan~which considers the latent-aspects of the claim to verify its truthfulness.
    \item In addition to classifying the claims,  \system~also provides evidence snippets highlighting the importance of both words and sentences relevant for classifying the claim using attention weights from the \sadhan~deep neural network model \cite{Mishra:2019:SHA:3341981.3344229}. 
\end{enumerate}

%% file: relatedwork.tex
\section{Related Work} 
\label{sec:relatedwork}

\begin{figure*}[ht!!!]
    \centering
    \includegraphics[scale=0.6]{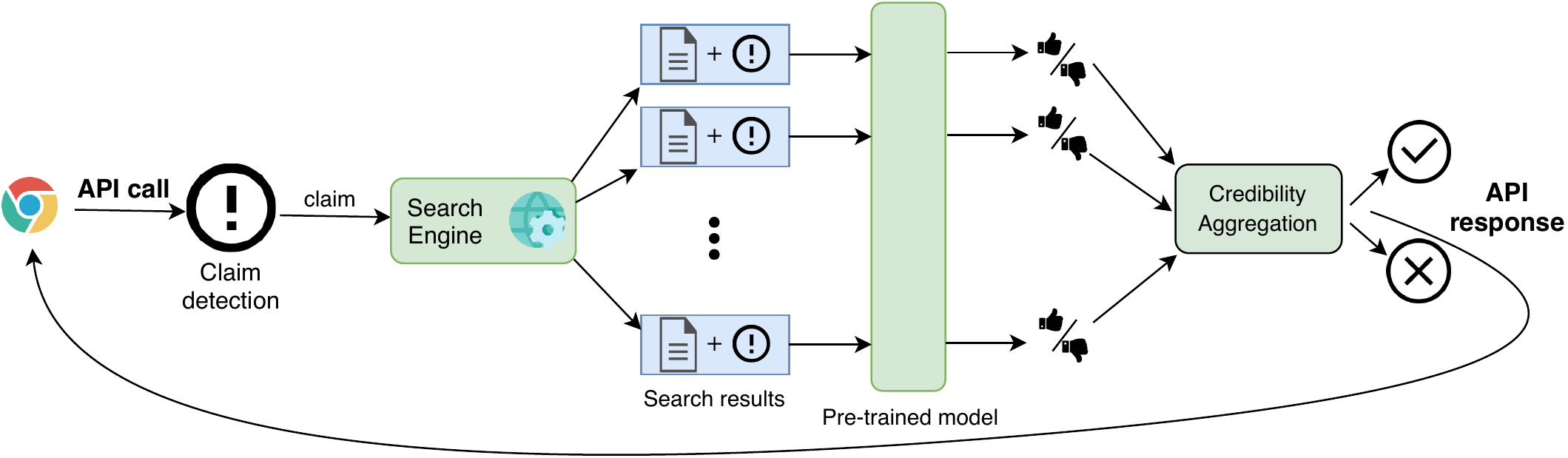}
    \caption{Block diagram of the server.}
    \label{fig:pipeline}
\end{figure*}

\begin{figure*}[ht!!!]

    \centering
\includegraphics[scale=0.25]{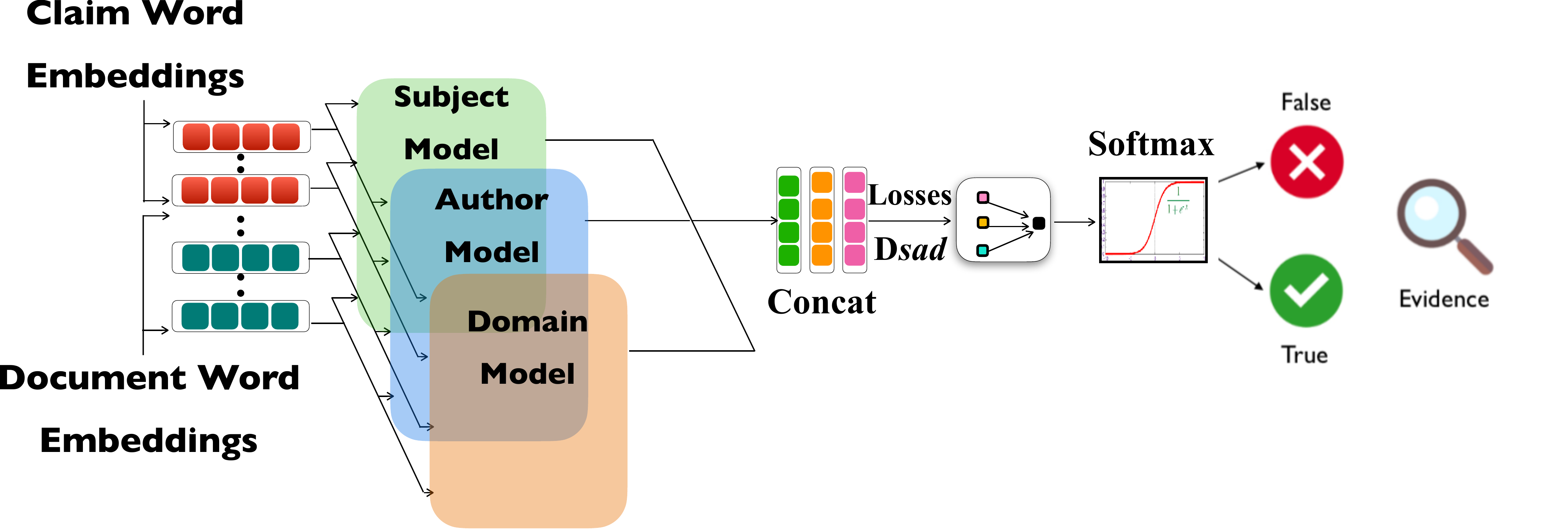}

    \caption{\sadhan~Model}
      \label{fig:SADHAN_model}

\end{figure*} 
Most fact-checking websites such as Snopes.com and Politifact.com perform manual fact check. Some automated fact-checking systems such as CredEye \cite{Popat:2018:CCL:3184558.3186967} are available. However, since CredEye only uses word-level attention, it can only highlight which words were used for classifying a claim. \system~on the other hand can provide evidence at both-word level and sentence-level. Moreover, \system~can provide evidence w.r.t each aspect such as subject, author and domain of the claim. FactMata is a commercial tool for automated fact-checking, there is no description of the detection algorithm. Moreover, they do not provide any evidence snippets. Grover\footnote{https://grover.allenai.org}\cite{zellers2019defending} is another solution which focuses on detecting neural generated fake news. To the best of our knowledge none of these systems are provided as a browser extension which allows users to fact-check without leaving the article they are reading. 

There are some browser extensions such as The Factual\footnote{\url{thefactual.com}}, Trusted Times\footnote{\url{trustedtimes.org}}, and FakerFact\footnote{\url{fakerfact.org}} which claim to support automated fact checking and they are listed in google chrome extension store. However, there is no research paper or documentation explaining the model they use.  Moreover, we could not find any system which can narrow down the claim within the article using fact-check worthiness detection and use that claim to detect fake news.

%% file: system.tex
\section{System Design}
\system~follows a client-server architecture and has a frontend and backend module. The frontend is a browser extension and the backend is a python Flask server.
\subsection{Frontend: Browser Extension}
We develop a browser extension which works with the popular Google Chrome browser. When the user invokes the fact checking by clicking on the browser extension, JavaScript modules are used to retrieve information and details from the web pages and send the query to the server. When the results are returned back from the server, another JavaScript module is invoked to display the results.

\subsection{Backend: Server}
\label{sec:server}
The server provides a RESTful API for the browser extension. The browser extension sends the URL or claim text chosen by the user to the server. The server then analyzes the claim text first by retrieving relevant articles from the Web via search engines such as Google and analyzes them by applying machine learning models and gives a prediction for the credibility of the claim.  A score indicating how credible the claim is based on the evidence found is sent back to the browser. In this section, we explain different parts of the server. The overall block diagram of the server can be seen in Figure \ref{fig:pipeline}.

\paragraph{Querying the Web:}
Given a claim text, we use Google API to retrieve the top-10 relevant web pages. We use the claim text as the query without quotes. Before passing the text to the neural network for credibility prediction, we preprocess the text to tokenize, extract publication date, authors and summary etc using a python library \textit{Newspaper3k}\footnote{\url{https://newspaper.readthedocs.io/}}. Since not all parts of the news article are important to classify the claim, we filter the articles with relevant snippets using cosine similarity (inspired by \cite{popat2018declare}). Then we select all the snippets above 0.75 similarity score for fact checking.

	 \begin{table}[t!!!]
		\centering
		\caption{Comparison of \sadhan~with $\declare$ models for False claim detection on Snopes and PolitiFact datasets. }
		\scalebox{0.75}{
			\begin{tabular}{p{1cm}p{2cm}p{1.5cm}p{1.5cm}p{1.5cm}p{1.5cm}}
				\hline
				\bf Data &\bf Model & \bf True Acc. & \bf False Acc. & \bf  Macro F1  & \bf AUC  \\
				\hline
				& \declare & 68.18 & 	66.01 &	67.10 &	72.93 \\

					\politifact                      & \sadhan &\textbf{68.37}& \textbf{78.23}&\textbf{75.69}&\textbf{77.43}\\
				\midrule

					                 & \declare & 60.16 & 	80.78 &	70.47 &	80.80 \\
				                
				\snopes                   & \dhan &\textbf{79.47}&\textbf{84.26}&\textbf{80.09}& \textbf{85.65}\\
\hline
		\end{tabular}
		}
		\vspace{-10pt}
		\label{tab:results_1}
	\end{table}

	\begin{figure*}[t!]
    \centering
    \subfigure[Popup]{\includegraphics[width=0.3\textwidth, frame,  page=1]{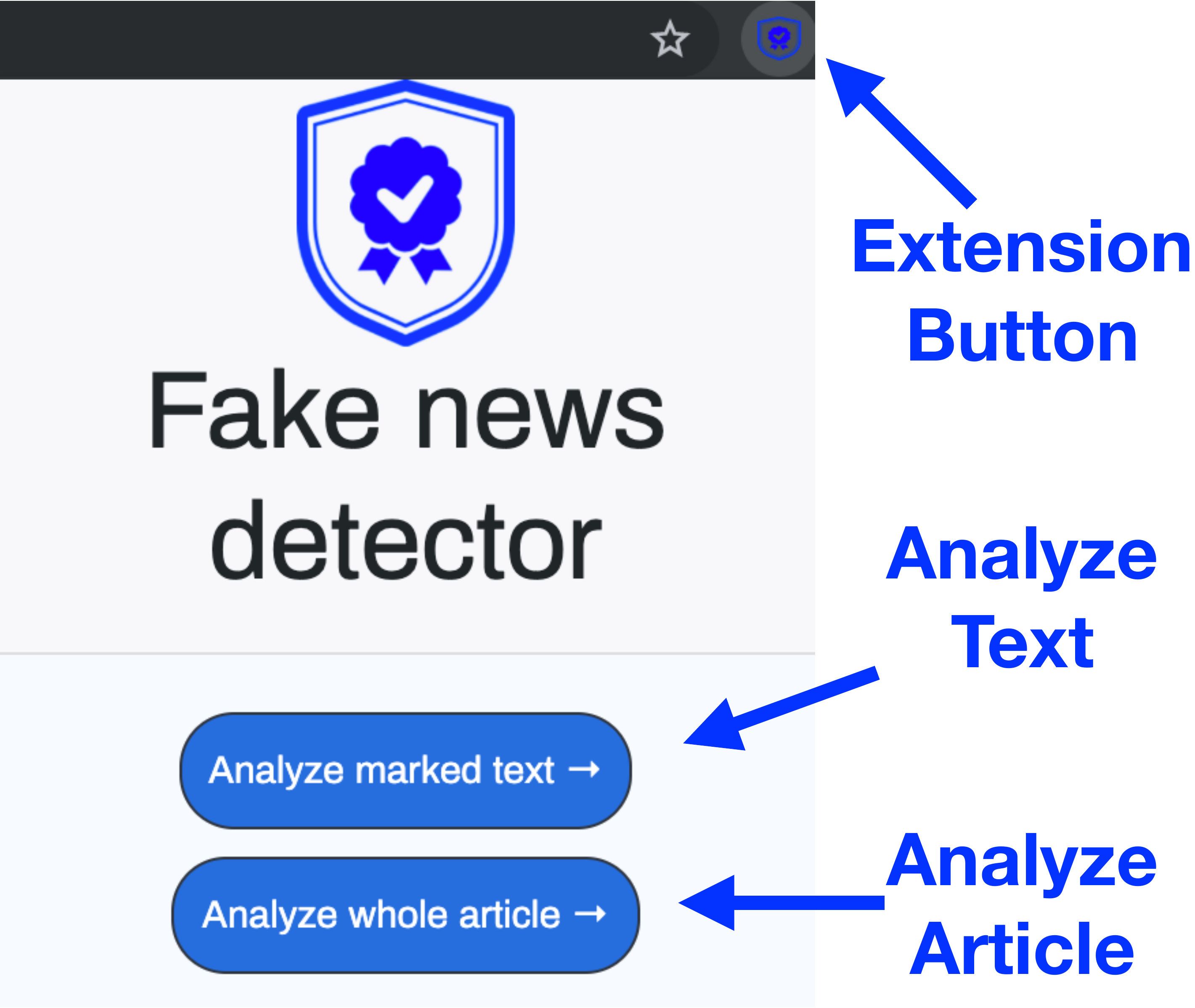}\label{fig:popup}}
    \subfigure[Result]{
    \includegraphics[width=0.6\textwidth , frame,  page=2]{annotations.pdf}
    \label{fig:resultpage}}
    \centering
    \subfigure[User feedback for the  result]{\includegraphics[width=0.45\textwidth, frame,  page=3]{annotations.pdf}\label{fig:feedback}}
    \subfigure[Claim extraction]{
    \includegraphics[width=0.4\textwidth, frame,  page=4]{annotations.pdf}
    \label{fig:claimfilter}}
    
    \caption{Demonstration Snapshots}
\end{figure*}

\paragraph{\sadhan~Model:}
In this demo, for the classification of fake news articles and false claims we use a deep neural network coined \sadhan~\cite{Mishra:2019:SHA:3341981.3344229}. \sadhan~model  uses hierarchical neural attention mechanism \cite{yang2016hierarchical} for learning the representations for both claim text and the evidence news article both at word level and sentence level. As shown in Figure \ref{fig:SADHAN_model}, \sadhan~takes claim text and a evidence document embeddings as input. Optionally, \sadhan~can also take latent aspects such as `author', `topic' and the `domain' etc into account to guide the attention. The aspect attribute vector used in computation of attention at both the word and sentence level comes from latent aspect embeddings for which weights are trained jointly in the model using
corresponding aspect attentions. As shown in Table \ref{tab:results_1}, \sadhan~outperforms powerful baselines which uses word-level attention such as \declare~\cite{popat2018declare}. For more details and performance evaluation of \sadhan~see \cite{Mishra:2019:SHA:3341981.3344229}.

\paragraph{Claim Detection:}

Since not all sentences in the articles are worthy of a fact check, we train a classifier and use it for detecting the claim check worthy sentences. We use ULMFiT, a language model fine-tuning technique \cite{howard2018universal} and use a model inspired by Averaged-SGD-LSTM \cite{DBLP:journals/corr/abs-1708-02182} to train our classifier.  The model is trained with a dataset with 9069 labeled sentences (4094 from a  presidential debate dataset \footnote{\url{https://github.com/apepa/claim-rank/tree/master/data}} and 4975 from the Politifact dataset \footnote{\url{politifact.com}}. We combined these two datasets and together the dataset has 4666 with label ``claim'' and 4193 with label ``non-claim''. We performed 5-fold cross validation and got a precision of 0.913, a recall of 0.937 and F1-score (micro) of 0.920. We use the softmax value of the model as a claim-check worthiness score for the given sentence.

%% file: demo.tex
\section{Demonstration}

\begin{figure*}[ht!!!]

     	\centering  
	\includegraphics[scale=0.45]{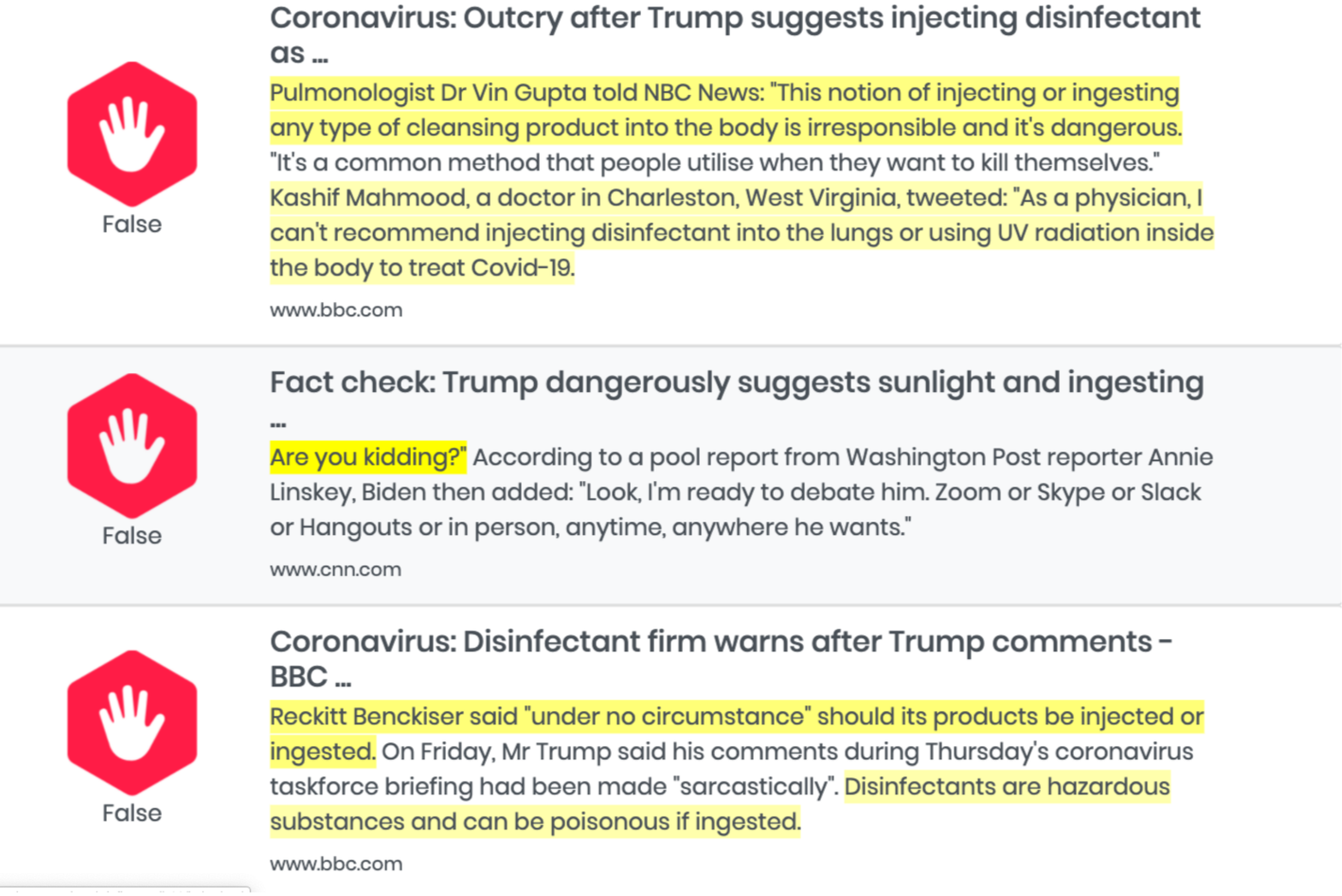}
\vspace*{-10pt}
     	\caption{ Evidence visualization for the claim ``Covid-19 can be cured by ingesting disinfectants'' using the attention weights}
     	\label{fig:attention1}

    \end{figure*}

The screen recording of the demonstration can be found here\footnote{\url{https://www.youtube.com/watch?v=LjaqH5JogGo}}. When the user invokes \system,~a popup is launched where the user can choose with what method they want to analyze the article. The user can then choose one of the two options, as shown in Figure~\ref{fig:popup}. When the ``Analyze marked text'' is chosen the selected text is used as the claim and sent to the server, which then runs the series of web page extraction, NLP and classification explained in Section \ref{sec:server}. The result from the \sadhan~model is displayed in the same popup window (See Figure \ref{fig:resultpage}). The user can also choose to see the evidence by clicking on the ``evidence'' button, which then extracts the evidence snippet according to the attention mechanism of \sadhan~model.

Users can also give a feedback if the model makes a mistake (Figure \ref{fig:feedback}) which in-turn could be potentially used to improve the classifier or evaluate the performance on the live data. When ``Analyze the whole article'' is clicked, another popup shown in Figure \ref{fig:claimfilter} is launched. \system~automatically analyzes the whole article and fact checks the top scored claim using \sadhan~model. The user can also explore other identified claims in the article by setting the claim score threshold and the top-k sentences. The user can also provide feedback on claim score prediction by our model.

When the user clicks on the evidence button, the user can also see the highlighted sentences based on the attention mechanism in \sadhan~model \cite{Mishra:2019:SHA:3341981.3344229}. The sentence-level attention weights are aggregated using word-level attention weights. This provides an intuitive understanding of the text the model considered as important for the classification. For example, in Figure \ref{fig:attention1},  for the claim ``Covid-19 can be cured by ingesting disinfectants'' the evidence is shown with highlighted sentences with contrast of the color proportional to the normalized aggregated word-level attention weights. 

The Chrome browser extension along with the instructions on how to install it can be found here\footnote{\url{http://factiverse.no/brenda/brenda_extension.zip}}.